\documentclass[twocolumn,showpacs,preprintnumbers,amsmath,amssymb]{revtex4}
\usepackage{graphicx}
\usepackage{dcolumn}
\usepackage{bm}

\begin{document}


\title{Analysis of the convergence of the $1/t$ and Wang-Landau algorithms in the calculation of multidimensional integrals.}

\author{R. E. Belardinelli}
  \email{rbelar@unsl.edu.ar}
\author{S. Manzi}
\email{smanzi@unsl.edu.ar}
\author{V. D. Pereyra}%
\email{vpereyra@unsl.edu.ar}

\affiliation{%
Departamento de F\'{i}sica, Instistuto Nacional de F\'{i}sica Aplicada, Universidad Nacional de San Luis, CONICET, Chacabuco 917,5700 San Luis, Argentina.
}%

\date{\today}

\begin{abstract}
In this communication, the convergence of the $1/t$ and Wang - Landau algorithms in the calculation of multidimensional numerical integrals is analyzed. Both simulation methods are applied to a wide variety of integrals without restrictions in one, two and higher dimensions. The errors between the exact and the calculated values of the integral are obtained and the efficiency and accuracy of the methods are determined by their dynamical behavior. The comparison between both methods and the simple sampling Monte Carlo method is also reported. It is observed that the time dependence of the errors calculated with $1/t$ algorithm goes as $N^{-1/2}$ (with N the MC trials) in quantitative agreement with the simple sampling Monte Carlo method. It is also showed that the error for the Wang - Landau algorithm saturates in time evidencing the non-convergence of the methods. The sources for the error are also determined.
\end{abstract}

\pacs{02.60.Jh; 05.10.Ln; 02.70.Uu}

\maketitle

\section{\label{sec:level1}INTRODUCTION}

It is well-known that Wang-Landau (WL) algorithm is one of the most refreshing variations of the Monte Carlo simulation methods introduced in the last time \cite{wang01}. Its effectiveness is based on the simplicity and versatility of the algorithm to calculate the density of state $g(E)$ with high accuracy (here $g(E)$ represents the number of all possible states or configurations for an energy level $E$ of a given physical system). In fact, on visiting states with energy $E$\textbf{} the running estimate $g_e(E)$  is multiplied by the refinement parameter $f>1$, which forces the system to visit less explored energy regions through the bias acceptance probability of min[1, $\frac{g_e(E_i)}{g_e(E_f)}$] for a move $E_i\rightarrow E_f$ (with $i$ and $f$ the initial and final states) and enables a fast performance compared to other flat energy histogram methods \cite{berg91}.

The algorithm has been successfully used in many problems of statistical physics, biophysics and others \cite{wang01, douarche03,troyer03,malakis04,brown05,schulz05A,reynal05,jayasri05,trebs05,ratore02,yan02,chopra06,shell02,yamaguchi01,okabe02,fasnacht04,Mastny05,carri05,calvo02,faller03,tsai07,vorontsov04,paulain06}. The method is based on independent random walks which are performed over adjacent overlapped energy regions, providing the density of states  \cite{wang01,schulz03B}. In that way, thermodynamic observables, including the free energy over a wide range of temperature, can be calculated with one single simulation.

There have been several papers in recent years dealing with improvements and sophisticated implementation of the WL iterative process in discrete and continuous systems  \cite{troyer03,jayasri05,shell02,paulain06,schulz03B,depablo03,shell03,zhou05,troster05,liang05,zhou06,kim06,berg07}.
However one of the most controversial point in the application of the WL and others variations of the algorithm is the saturation of the error between the calculated and the real $g(E)$. In fact, its approaches to a constant as a function of the MC time, for enough long time.

This problem was firstly evidenced by Q. Yang and J. J. de Pablo in reference \cite{depablo03}. On the other hand, other authors \cite{zhou05,lee06,sinha07,Earl05,Morozov07} have studied the accuracy, efficiency and convergence of WL algorithm. Some of them \cite{zhou05,Earl05,Morozov07} have demonstrated the convergence of WL algorithm by different arguments. Particularly, Zhou and Batt \cite{zhou05} have present a mathematical analysis of the WL algorithm. They give a proof of the convergence and the sources of errors for the WL algorithm and the strategies for improvement.

The saturation of the error and the convergence of the algorithm are certainly two contradictory results. Moreover, is a crucial breakpoint because the saturation of the error means the non convergence of the algorithm.
This polemic point has been treated in reference \cite{belardinelli07A,belardinelli07B}, where an analytical demonstration of the non convergence in the original version of the WL algorithm has been presented \cite{belardinelli07B}. Alternatively, the authors have deduced analytically the way to avoid the saturation of the error given an adequate form to the refinement parameter. In fact, those methods in which the refinement parameter vary lower faster than $1/t$ (with $t$ the Monte Carlo time) determine that the calculated density of states reaches a constant value for long times, therefore the error saturates. To overcome this limitation, they introduced a modified algorithm in which the refinement parameter is scaled down as $1/t$ instead of exponentially \cite{belardinelli07A}. This new algorithm allows the calculation of the density of states faster and more accurately than with the original WL algorithm due to the fact that the calculated density of states function approaches asymptotically the exact value. The $1/t$ algorithm has been successfully applied to several statistical system \cite{belardinelli07A,belardinelli07B} including the protein folding \cite{ojeda07}.

The non convergence of the original WL algorithm and other previous version, including N-fold way method \cite{schuz02,Malakis04}, seemed very difficult to believe. However, is interesting to emphasize that Landau and co-workers in Ref. \cite{landau04} suggest that $ln(f_{final})$ cannot be chosen arbitrarily small or the modified $ln[g(E)]$ will not differ from the unmodified one to within the number of digits in the double precision numbers used in the simulation. If this happens, the algorithm no longer converges to the true value, and the program may run forever. If $ln(f_{final})$ is within the double precision range is too small, the calculation might take excessively long to finish.

Although, the saturation of the error and consequently the non convergence of the WL algorithm have been demonstrated for a discrete system \cite{belardinelli07B}, namely the Ising Model. The mathematical arguments of the source of the error for the WL algorithm seem to be more general and can be extended to all algorithms which consider a refinement parameter that change, according to the flatness condition of the energy histogram, with a law that decreases faster than $1/t$. In all these cases, a saturation of the error for the calculation of the density of states and consequently the non convergence of the methods can be guaranteed.

Recently, Y. W. Li et al. \cite{li07} report a new application of the well-known Wang-Landau algorithm sampling to the simplest continuous systems, namely the numerical integration. The basic idea of this new application of the WL algorithm is to establish a parallel between the density of states g(E) and the distribution g(y). Here g(y) represents the fraction of the integration domain ($[a,b]$ in one-dimension) that lies within a certain interval $[y,y+dy]$.

This idea was proposed firstly by Tr\"oter and Dellago \cite{troster05}. The authors adapted a Wang-Landau sampling scheme to the problem of numerical integration as an application of their self-adaptive range Wang-Landau algorithm. In their approach, the integrand $y(x)$ is expressed in terms of a "Boltzmann factor" $e^{-\phi(x)}$ with $\phi(x)=ln(y(x))$ and $k_BT=1$ and a random walk in this so defined energy space is performed. Simultaneously, Liang \cite{liang05} was developed a generalization of the WL algorithm to continuous systems. This methods was used to the MC integration and MC optimization. The scheme facilitates the numerical integration in case of sharply peaked functions. However, both methods exhibit a severe restriction, namely $y(x)>0$.

In principle, the WL method of integration \cite{li07} presents various advantages on the conventional MC integration scheme. Specifically, these are: (i) it provides a procedure for the numerical integration of sharply-peaked or ill-behaved integrand which is difficult to be dealt with conventional MC methods. (ii) It can be used for integrands with negative values. The correspondence between the density of states $g(E)$ (physical system) and $g(y)$ (integration) does not require a functional form as is proposed in reference \cite{troster05,liang05}. (iii) It is not necessary to known the boundaries of the integrand such as the global minimum and maximum of the function $y(x)$ within the integration domain. iv) It seems that the flatness criterion $p$ and the bin width $dy$ provide two adjustable parameters which allow the control of the accuracy of the numerical estimate. The bin width seems to be the predominant parameter for attaining reasonable estimates.

Although the accuracy of the WL algorithm is worse than the simple sampling MC, at least for one and two dimension, as is shown in ref. \cite{li07}, its potential rather comes up for ill-behaved integrals and for higher-dimensional integration problems in general since the random walk remains one-dimensional. However, the problem of convergence of the algorithm is of a crucial importance in the accuracy of the numerical calculation for certain integrals.

Moreover, the multidimensional numerical integration seems to be the more adequate test laboratory to prove the convergence of the WL and any other algorithms. There are two main reasons for that, the first one is associate to the fact that, as a difference with the statistical models where the real $g(E)$ is only known in few discrete cases, for numerical integration real $g(y)$ can be calculated exactly for some one-to-one functions. On the other hand, for those well-behaved functions, it is always possible to calculate the numerical integral using the simple sample Monte Carlo simulation.

Beside that, for continuous systems, there are few simulations where comparison has been made with the exact density of states  \cite{Mukhopadhyay08}. The reason for that may be attributed to the non-availability of results of exact calculation for any non-trivial system having a continuous energy spectrum.

In order to prove such arguments, an analysis of the convergence and saturation of the error for both, the WL and $1/t$ algorithms is presented. The study is developed in the framework of numerical calculation of multidimensional integrals.

It is necessary to emphasize that our objective is to discriminate the source of errors in both algorithms by using the dynamical behavior of the error, instead of presenting a new method of multidimensional integration.

The outline of the present paper is as follows: in Section 2, it is introduced the  $1/t$ algorithm adapted to the numerical integration. In Section 3, several examples which include up to six-dimensions numerical integrals are introduced. The sources of the errors in both methods are discussed. Finally, the conclusions are given in Section 4.

\section{$1/t$ ALGORITHM AND THE NUMERICAL INTEGRATION}

It is well-known that Monte Carlo methods are an efficient alternative to calculate numerical integrals in higher dimensions. Technically, Monte Carlo integration is numerical quadrature using pseudorandom numbers \cite{kooning85,goul06,Teukosky92,landau05}. That is, Monte Carlo integration methods are algorithms for the approximate evaluation of definite integrals, usually multidimensional ones. The usual algorithms evaluate the integrand at a regular grid. Monte Carlo methods, however, randomly choose the points at which the integrand is evaluated.

The first versions of the method are rather limited, for instance, simple sampling Monte Carlo integration suffers from slow convergence requiring a large amount of sampling to reduce the statistical error. However, convergence is even not always assured.

Several variations of the classical MC integration have been introduced in order to improve the performance of the method.

For instance, the importance sampling Monte Carlo method \cite{kooning85} reduces considerably the statistical error by sampling points which are generated according to a probability distribution $p(x)$ and the flattened ratio $y(x)/p(x)$ is integrated instead of the original integrand $y(x)$. The main limitation of this procedure arises because the weighting probability has to be positive and normalized to unity in the integration domain \cite{goul06}. On the other hand, importance sampling methods may even converge to incorrect values if a bad weighting function is chosen; however, such errors are not readily detected. Moreover, conventional MC integration methods fail or are less efficient in case of sharply peaked or ill-behaved functions on multidimensional domains.

Other methods have been introduced in the past to improve the importance sampling Monte Carlo integration, as for example the VEGAS algorithm \cite{lepage78}. It samples points from the probability distribution described by the function $|f|$ , so that the points are concentrated in the regions that make the largest contribution to the integral.

In this section, the $1/t$ algorithm is adapted to the numerical calculation of multidimensional integrals. The basic idea is as follows: To evaluate the definite integral $\int_a^b y(x) dx$ is necessary to determine the proportion of integration domain that lies within a certain interval $[y, y+dy]$, i.e. the measure $\lbrace x\mid x \in [a, b], y \leq y(x) \leq y+dy \rbrace$. The distribution $g(y)$ can be generated measuring this fraction. As it point out below, this quantity is a direct analogy to the density of states $g(E)$ of a physical system. Provided that the lower bound $y_{min}$ and the upper bound $y_{max}$ of the integral are known the integral can then be approximated by
\begin{eqnarray}
I=\int_a^b y(x)dx \approx \sum_{y_{min}}^{y_{max}}g(y)y
\end{eqnarray}

To build the distribution $g(y)$, the interval $[y_{max}-y_{min}]$ is divided in $L=[y_{max}-y_{min}]/dy$ segment. The MC time is defined as $t=N/L$, where $N$ is the number of Monte Carlo trials. In what follows all the quantities will be related to the numbers of MC trials $N$, in order to compare the algorithms.

In practice, the relation $S(y)=ln[g(y)]$ is generally used, in order to fit all possible values of $g(y)$ into double precision numbers.

The algorithm is as follows:

i) Choose a value of $x_i$ at random and the corresponding value of $y_i$ is calculated; then set $S(y)=ln[g(y)]=0$ for all value of $y$, $F_o=1$ and fix $F_{final}$ or equivalently $t_{final}=1/F_{final}$.

ii) A value $x_f$ is also chosen at random and the system changes from $y_i$ to $y_f$ according to the probability given by

\begin{eqnarray}
P(y_i \rightarrow y_f)=min  \left \{ 1,\frac{g(y_{i})}{g(y_{f})} \right \}\nonumber\\=min \left \{1,e^{[S(y_{i})-S(y_{f})]} \right \}
\end{eqnarray}

iii) Increment $S(y)\rightarrow S(y)+F_k$.

iv) After some fixed sweeps (i.e., 1000 MC time) check that all the sites "y" corresponding to the same $F_k$ will be visited by the random walker at least one time, then refine $F_k=F_k/2$.

v) If $F_{k+1}\leq 1/t=L/N$ then do $F_{k+1}=F(t)=1/t=L/N$. In what follows $F(t)$ is updated at each MC time. The step iv) is not used for the rest of the experiment.

vi) If $t>t_{final}$ ($F(t)< F_{final}$) then the process is stopped. Otherwise go to ii).

Note that, with the exception of $S(y)$, the $1/t$ algorithm does not use any auxiliary histogram in the calculation of distribution $g(y)$.

At short times, the random walker must visit a given site $y$ at least one time with the same $F_k$. As soon as the refinement parameter takes $F=1/t$ functionality, it goes down independently of the number of times that the site $y$ is visited by the random walker.

As in the original WL procedure \cite{wang01,li07}, our algorithm provides only a relative distribution function $g(y)$; however, in order to evaluate the integral, $g(y)$ needs to be normalized appropriately. In one-dimension, the normalized $g_{norm}(y)$ is obtained by

\begin{eqnarray}
g_{norm}(y) = \frac{(b-a)g(y)}{\sum_{y_{min}}^{y_{max}}g(y)}\
\end{eqnarray}

The lower and the upper bounds, $y_{min}$ and $y_{max}$, respectively, of the integrand $y(x)$, as well as $y$ values that cannot be reached within the integration domain, have to be determined beforehand in order to ensure the feasibility of the procedure. As is discussed previously in reference \cite{troster05}, one possible way to find the valid range in y-space is to carry out an initial "domain sampling run" with $F=1$, before starting the actual iteration process. The $1/t$ algorithm can be easily generalized to higher dimensions as the original WL method \cite{li07}.

The main objective of the present work is the comparison of the dynamical behavior of the error using the $1/t$, Wang Landau (WL) and  Simple Sampling (SS) algorithms. In what follows, the same nomenclature introduced in reference \cite{li07} is used . Therefore the fractional accuracy is given by

\begin{eqnarray}
a_f(N) =  \left |  \frac{I_{MC}(N)-I_{exact}}{I_{exact} }\ \right  |
\end{eqnarray}
where $I_{MC}(N)$ denotes the numerical estimate from Monte Carlo procedure and $I_{exact}$ is the exact value of the integral.

\section{DISCUSSION OF THE COVERGENCE. A COMPARISON BETWEEN ALGORITHMS}

In this section,  numerical integrals are calculated by using the three algorithms ($1/t$, WL and SS). The first two integrals given in ref.\cite{li07} are,

\begin{eqnarray}
I_{1D}=\int_{-2}^{2}(x^5-4x^3+x^2-x)\sin{(4x)}dx,
\end{eqnarray}
in one dimension, which its exact value is $I_{1D}=1.63564436296...$, and

\begin{eqnarray}
I_{2D}=\int_{-1}^{1}\int_{-1}^{1}(x_1^6-x_1x_2^3+x_1^2x_2+2x_1)\nonumber\\
\times\sin{(4x_1+1)}\cos{(4x_2)}dx_1dx_2,
\end{eqnarray}
in two-dimension, which its exact values is $I_{2D}=-0.01797992646...$.

Then,it is evaluated the next one-dimensional integral,

\begin{eqnarray}
I_{\pi}=\frac{1}{4}\int_{0}^{1}\sqrt{1-x^2}dx,
\end{eqnarray}

which in the first quadrant, its exact value is $I_{\pi}=\pi$ .

Next, it is evaluated the following multidimensional integrals,

\begin{eqnarray}
I_{nD}=\int_{0}^{1}\int_{0}^{1}...\int_{0}^{1}\prod_{i=1}^{i=n}\cos{(ix_i)}dx_1dx_2...dx_n,
\end{eqnarray}
which their exact values can be easily obtain as,
\begin{eqnarray}
I_{nD}=\prod_{i=1}^{i=n}\frac{\sin{(ix_i)}}{i}
\end{eqnarray}

It is necessary to emphasize that those integrals given above have no particular physical or mathematical significance. However, their are very useful to compare the convergence of the three MC algorithms.

Next, the results are shown. All the error estimates are obtained from $100$ independent simulations.

\begin{figure}
\includegraphics[scale=0.85]{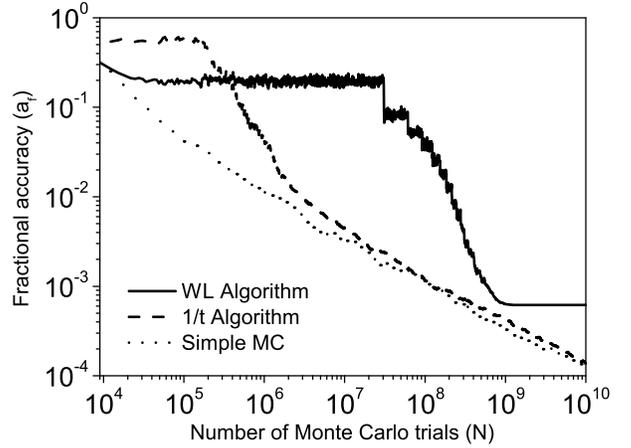}
\caption{\label{fig:epsart} Dynamical behavior of  fractional accuracy  $a_f$ for one-dimensional integral given in eq.(5) calculated by means of Wang-Landau, 1/t and simple sampling Monte Carlo integration. The WL calculations have been made with a flatness criterion $ p=0.9$. In both cases, the WL and $1/t$ algorithms the bin width is $dy=0.005$. The quantities are obtained averaging over 100 independent samples.}
\end{figure}

\begin{figure}
\includegraphics[scale=0.85]{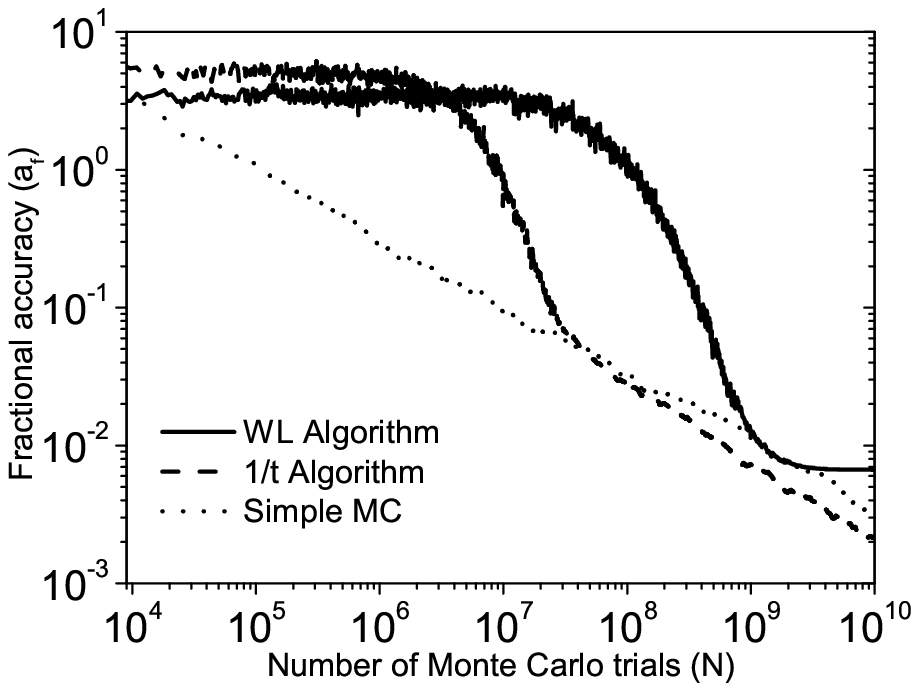}
\caption{\label{fig:epsart} Dynamical behavior of  fractional accuracy  $a_f$ for two-dimensional integral given in eq.(6) calculated by means of Wang-Landau, 1/t and simple sampling Monte Carlo integration. The WL calculations have been made with a flatness criterion $ p=0.9$. In both cases, the WL and $1/t$ algorithms the bin width is $dy=0.005$. The quantities are obtained averaging over 100 independent samples.}
\end{figure}

In the Figure 1, the fractional accuracy for the one-dimensional integral $I_{1D}$ given in eq.(5), as a function of the MC trials calculated by the three algorithms is shown. The WL calculations have been made controlling the histogram every $10000$ MC with a flatness criterion $ p=0.9$. In both cases, the WL and $1/t$ algorithms the bin width is $dy=0.005$ and $L=3066$. Clearly, the fractional accuracy for the WL algorithm saturates for $N\approx 10^9$, while the $1/t$ and simple sampling  calculations of the error are in close agreement with a behavior given by $a_f \propto 1/\sqrt{N}$, as is expected.

In Figure 2, the fractional accuracy for the two-dimensional integral $I_{2D}$ given in eq.(6), as a function of the MC trials calculated by the three algorithms is shown. In both cases, the WL and $1/t$ algorithms the bin width is $dy=0.005$ and $L=1213$. The flatness criterion for WL algorithm is the same used in Figure 1. The behavior of the error is similar to the one-dimensional case presented in Figure 1.

As one can observed in both cases the error for WL saturates, then $a_f$ does not scale as $N^{-1/2}$.

\begin{figure}
\includegraphics[scale=0.85]{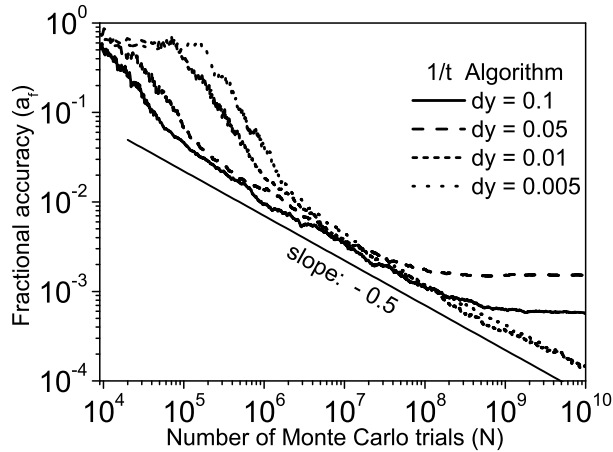}
\caption{\label{fig:epsart} Dynamical behavior of fractional accuracy $a_f$ for different values of bin width $dy$ by means of 1/t algorithm. The same integral and condition described in Figure 1 is used.}
\end{figure}

\begin{figure}
\includegraphics[scale=0.85]{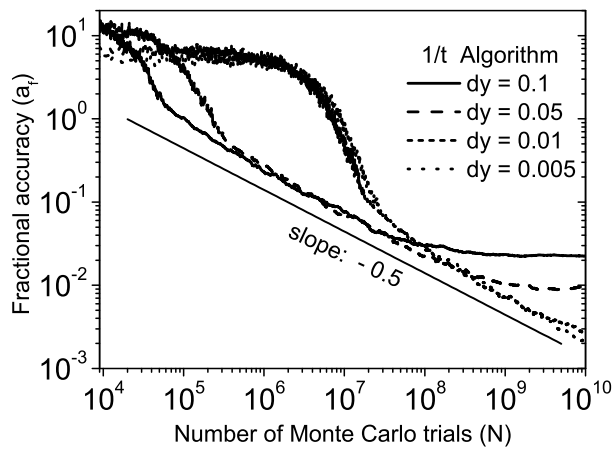}
\caption{\label{fig:epsart}Dynamical behavior of fractional accuracy $a_f$ for different values of bin width $dy$ by means of 1/t algorithm. The same integral and condition described in Figure 2 is used.}
\end{figure}

The bin width certainly introduces a systematic error in all the algorithms that use the distribution function $g(y)$ as strategy to calculate numerical integrals. In Figure 3, the effect of $dy$ in the calculation of the error for the $1/t$ algorithm (for the same integral and condition given in Figure 1), is shown. The dependence of the error with the bin width will be determined by the characteristic of the function $y(x)$. In fact, one can expect that for smaller $dy\neq 0$ the saturation will occur at longer times. However, that is not generally valid, as is observed in the figure where for $dy=0.05$ the error saturates before than for $dy=0.1$.

In Figure 4, the effect of the bin width $dy$ on the behavior of $a_f$ for the integral referenced in Figure 2 is shown. In this case the smaller values of $dy$ lead to small errors. As a difference with the WL algorithm, the numbers of Monte Carlo trials is an input parameter for the $1/t$ and simple sampling algorithms.

To make a deep analysis of the effect of the bin width in behavior of the error in the $1/t$ algorithm, let us consider the integral given in eq.(7), which its exact value is $I_{\pi}=\pi$. For the definition of the distribution function and considering that $f(y)$ is one-to-one function, it is possible to obtain the exact value of $g(y)$ as,
\begin{eqnarray}
g_{ex}(y)=\sqrt{1-(y+dy)^2}-\sqrt{1-y^2}
\end{eqnarray}
which is valid for all $ dy \neq 0$. Then, for a given value of $dy$, the corresponding "exact" value of the integral $I_{\pi}(dy)_{ex}$ can be obtained. Let us define a new fractional accuracy related to the value of $dy$ as
\begin{eqnarray}
\bar a_f(N,dy) = \left | \frac{I_{MC}(N)-I_{\pi}(dy)_{ex}}{I_{\pi}(dy)_{ex}}\ \right |
\end{eqnarray}

Let us analyze the effect of the bin width in both WL and $1/t$ algorithms by calculating the integral given in eq.(7) and using the fractional accuracies given in eq.(4) and eq.(11).

\begin{figure}
\includegraphics[scale=0.85]{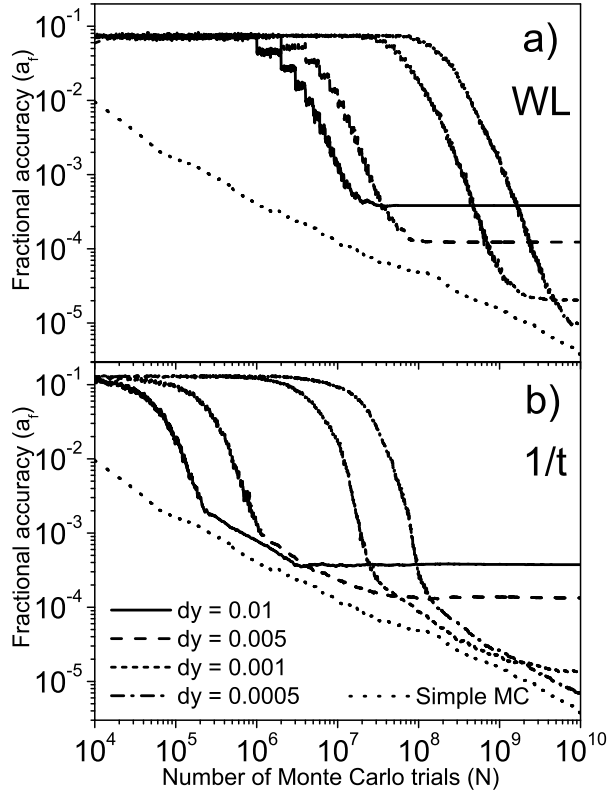}
\caption{\label{fig:epsart} Dynamical behavior of  fractional accuracy  $a_f$ for one-dimensional integral given in eq.(7) calculated by means of a) Wang-Landau, and b) 1/t algorithms for different values of bin width $dy$. The WL calculations have been made with a flatness criterion $ p=0.9$. In both cases, the results are compared with the simple sampling Monte Carlo integration. All the quantities are obtained averaging over 100 independent samples.}
\end{figure}

\begin{figure}
\includegraphics[scale=0.85]{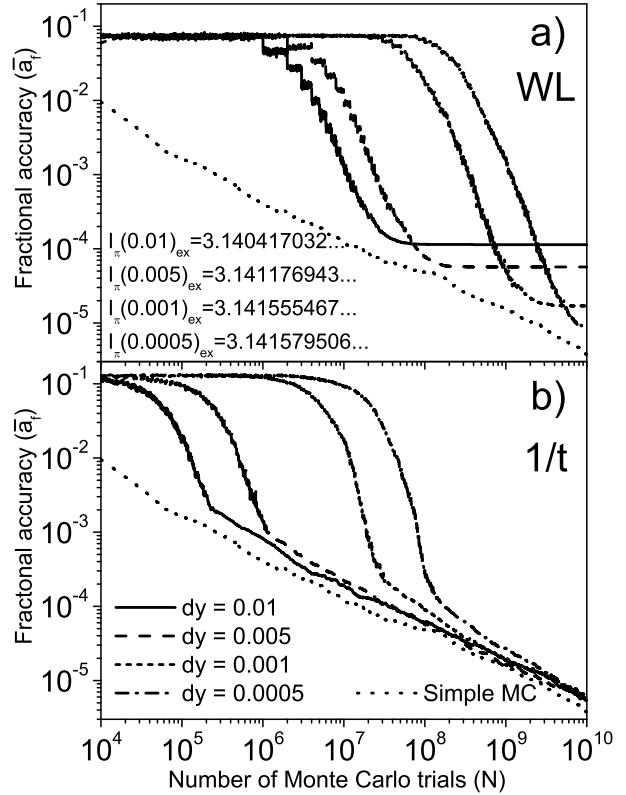}
\caption{\label{fig:epsart} Dynamical behavior of fractional accuracy $\bar a_f$ for different values of bin width $dy$, for the same integral and condition explained in Figure 5. In the inset of Figure a), the exact value of $I_{\pi}(dy)_{ex}$ is expressed. }
\end{figure}

In Figure 5 a) and b) it is shown the fractional accuracy defined in eq. (4) versus the MC trials for different values of the bin width $dy$ for the WL and $1/t$, respectively.
The error saturates in both cases as is expected. However, in the next figures, it is demonstrated that the sources of the saturation in both cases obey different causes. In fact, to avoid the effect of $dy$ on the saturation of the error, in Figure 6 a) and 6 b) it is shown the fractional accuracy defined in eq.(11).  Here for each value of $dy$ one obtain the exact value of $I_{\pi}(dy)_{ex}$. While for the WL results the error still saturates (Fig. 6 a) in $1/t$ calculations (Fig. 6 b) one can observe that no saturation occurs. This is a clear evidence of the convergence of the $1/t$ in all discrete systems.

Therefore, one can conclude that the source of saturation in $1/t$ algorithm is exclusively due to $dy$. In fact, for a given $dy$ the error approaches asymptotically to exact value of integral $I_{\pi}(dy)_{ex}$ which is different of $\pi$, as one can see in the inset of Figure 6 a). However, for WL algorithm the source of the error saturation are intrinsically associate to the nature of the algorithm, namely the decay of refinement parameter.

Finally, let us describe the behavior of both algorithms in the calculation of higher dimension integrals.
\begin{figure}
\includegraphics[scale=0.9]{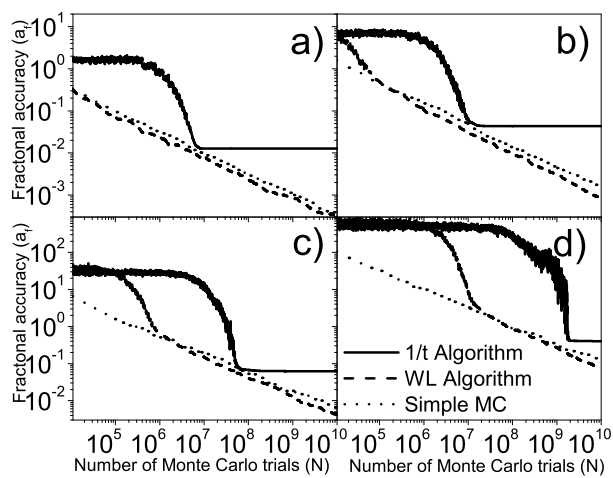}
\caption{\label{fig:epsart} Dynamical behavior of  fractional accuracy  $a_f$ for n-dimensional integral (8) with a) $n=3$, b) $n=4$, c) $n=5$, and d) $n=6$,  using Wang-Landau, 1/t and simple sampling Monte Carlo integration. The WL calculations have been made with a flatness criterion $ p=0.9$. In both cases, the WL and $1/t$ algorithms the bin width is $dy=0.05$. The quantities are obtained averaging over 100 independent samples.}
\end{figure}

In Figure 7 a) to d), it is plotted the integrals given in eq.(8) with $n=3,4,5,6$. The numerical calculations have been made using the three algorithms describe in the paper.
In all case $dy=0.05$, $L=40$.
As one can observe in all case the error for the WL algorithm saturates demonstrating that the non convergence of the method is independent of the dimension. On the other hand, the results obtained by using the $1/t$ algorithm is in close agreement with simple sampling.

The numerical estimates of statistical error for all the integrals are given in Table I, using the three algorithms described in the paper. The number of final Monte Carlo trials per run is $N_{final}=10^{10}$. All the WL calculations have been made with a flatness criterion $ p=0.9$. The values of the estimates showed in table I confirm that $1/t$ is more accurate than WL algorithm for all the integrals in any dimension.

\begin{table*}
\caption{\label{tab:table3} Numerical estimates of integrals calculated by using the Wang-Landau, 1/t and simple sampling Monte Carlo. The number of final Monte Carlo trials per run is $N_{final}=10^{10}$. All the WL calculations have been made with a flatness criterion $ p=0.9$. Results and error estimates are obtained from 100 independent simulations.}
\begin{ruledtabular}
\begin{tabular}{crrrr}
 Integral &  WL Algorithm  & 1/t Algorithm & Simple MC & exact \\ \hline
 $I_{1D}$&1.635580(285)\footnotemark[1] &1.635617(27)\footnotemark[1]&  1.635752(23)  &  1.63564436296...\\
 $I_{2D}$& -0.0179671(152)\footnotemark[1] & -0.0179790(51)\footnotemark[1] & -0.0179841(68)& -0.01797992646...\\
 $I_{\pi}$ &  3.1415799(35)\footnotemark[2] &  3.1415819(23) \footnotemark[2] &  3.1415920(15) &  3.14159265358...\\
 $I_{3D}$  &  0.01801608(2989) \footnotemark[3] &  0.01799079(46)\footnotemark[3] &  0.01799666(78)&  0.01799626791...\\
 $I_{4D}$  & -0.00339585(1798) \footnotemark[3] & -0.00340505(54) \footnotemark[3] & -0.00340506(55)& -0.00340490511...\\
 $I_{5D}$  &  0.00065747(505) \footnotemark[3] & 0.00065298(47)\footnotemark[3] &  0.00065225(49) &  0.00065300923...\\
 $I_{6D}$  & -0.000031919(3288)\footnotemark[3] & -0.00003003(15)\footnotemark[3] & -0.00003110(38)& -0.00003041015...\\

\end{tabular}
\end{ruledtabular}
\footnotetext[1]{The bin width is $dy=0.005$.}
 \footnotetext[2]{The bin width is $dy=0.0005$.}
 \footnotetext[3]{The bin width is $dy=0.05$.}

\end{table*}

\section{CONCLUSIONS}

In this work the numerical calculations of multidimensional integrals are used to analyzed the convergence of $1/t$ and WL algorithms. The numerical integration is an excellent laboratory to prove the convergence of the algorithms for different reasons:
i) in many cases the integrals can be solve analytically, then it can be easily checked the dynamical behavior of the error; ii) for some one-to-one functions, one can easily obtain the exact expression for the distribution function, $g(y)$. This is an advantage over the physical systems, where only in few cases the exact density of states is known, particularly for continuous system; ii) the  initial and final states are not correlate with their neighborhood, namely, a given initial state can be changed to any other final state in the integration domain of  $y(x)$. In this sense, if a given algorithm can not converge appropriately in the calculation of a numerical integrals, it will be more difficult to do it in those physical systems where the initial and final states are strongly correlated.

Although the flatness criterion $p$ and the bin width $dy$ provide a source of saturation of the error in the WL algorithm. The main reason of error saturation is that the refinement parameter is scaled down exponentially instead of a power law.

On the other hand, it is shown that the behavior of the error in the WL algorithm, for a single value of the parameters $dy$ and $p$, does not follows the $1/\sqrt{N}$ at any time.

Alternatively, the convergence of $1/t$ algorithm is analyzed through the calculation of multidimensional integrals.

To obtain the density of states function (physical systems) or the distribution function (integral calculation) in a continuous system, it is necessary a grid discretization. This introduces a systematic error in the calculation of the observables which depend on the size of the unitary cell of the grid, namely the bin width.

For this reason, the error in the $1/t$ algorithm saturates as a function of $dy$. However, when the continuous model is approached by a discrete lattice and the corresponding value of the distribution function can be obtained exactly, the calculation approaches asymptotically to the exact value of the integral without error saturation.

On the other hand, the behavior of the error in $1/t$ algorithm is in close agreement with the simple sampling, following the $1/\sqrt{N}$ law. Moreover, the $1/t$ algorithm can be used as a reference in the calculation of the density of state in physical systems due that up to now there are no other method which can calculate the error below the limiting curve $1/\sqrt{N}$.

In summary, in the present paper it is shown a new evidence of the saturation of the error in the WL algorithm which implies the non-convergence of the method.

In contrast, the dynamical behavior of the $1/t$ algorithm is analyzed, concluding that the algorithm is always convergent in all discrete system. In the continuous model, the only source of error saturation is the grid discretization. Therefore $1/t$ algorithm is a more efficient, accurate and easy implemented to simulate the distribution function (density of state) in discrete and continuous systems without using any histogram.\\

\section{ACKNOWLEDGMENTS}

We thank Prof. G. Zgrablich for reading the manuscript. This work is partially supported by the CONICET (Argentina).

\end{document}